\documentclass[%
 aip,amsmath,amssymb,floatfix,
 reprint,%
]{revtex4-2}

\usepackage{graphicx}
\usepackage{dcolumn}
\usepackage{bm}%
\usepackage[bookmarks=true,
colorlinks=true,linkcolor=blue,citecolor=blue,urlcolor=blue]{hyperref}
\usepackage{comment}
\usepackage[dvipsnames]{xcolor}

\usepackage[utf8]{inputenc}
\usepackage[T1]{fontenc}
\usepackage{mathptmx}

\begin{document}

\preprint{APS/123-QED}

\title[]{
Magnetic impurities in thin films and 2D Ising superconductors
}

\author{David M\"{o}ckli}
\email{d.mockli@gmail.com}
\author{Menashe Haim}%
\author{Maxim Khodas}%
 
\affiliation{%
 The Racah Institute of Physics, The Hebrew University of Jerusalem, Jerusalem 9190401, Israel
}%

\date{\today}

\begin{abstract}

In the theory of dilute magnetic impurities in superconductors, the effect of all impurity spin-components is expressed via a single magnetic scattering rate $\Gamma_\mathrm{m}$. In a more realistic setting, magnetic impurities are anisotropic.
In this case, the spatial randomness of three spin-components of impurities gives rise to  
generally different scattering rates $\Gamma_i$ ($i=1,2,3$).
We explore the effects of anisotropic magnetic impurities on the in-plane critical field in 2D superconductors. 
We discuss singlet, triplet and parity-mixed order parameters allowed in systems without the inversion center.
Also, the addition of a small amount of magnetic impurities may cause singlet to triplet crossovers. 
In all cases, different components of impurity spin affect the magnetic field -- temperature phase diagram differently. 
We show that anisotropy of the magnetic impurities can serve as a probe of unconventional triplet or parity-mixed superconductivity.

\end{abstract}

\maketitle

\section{Introduction}


The study of the effect of non-magnetic and magnetic impurities on superconductors has been of great importance, as it  
made the distinction between isotropic and anisotropic superconductivity possible \cite{Gorkov,Balatsky2006}.
Non-magnetic (scalar) impurities do not affect the isotropic singlet $s$-wave order-parameter. In stark contrast, a magnetic scattering is very efficient in destroying $s$-wave superconductivity. 
The non-magnetic and magnetic impurities are equally detrimental to the critical temperature $T_\mathrm{c}$ of isotropic and anisotropic superconducting states \cite{Daams1981,Schachinger1981,Maekawa1987} as summarized in Table \ref{tab:intro}.

The studies of magnetic impurities in superconductors date back to BCS theory \cite{Suhl1959b}. The traditional theories consider the magnetic exchange interaction $\boldsymbol{\sigma}\cdot \mathbf{S}$ between the spin of the conduction electrons $\boldsymbol{\sigma}$ and impurity spins $\mathbf{S}$ \cite{Brink1965,Fulde1966a,Kim1994a,Gorkov1964_rusinov,Gorkov}. 
When the impurities are dilute their spins are randomly oriented and uncorrelated \cite{Pint1989}.
An averaging procedure over all magnetic impurity sites yields a single magnetic scattering rate $\Gamma_\mathrm{m} \sim \overline{\mathbf{S}^2}  =S(S+1)$, where $S$ is the total impurity spin \cite{Rickayzen2013}. 
In this case, all $S_i$ ($i=1,2,3$) components contribute equally to $\Gamma_\mathrm{m}$, $\overline{ S_i^2} =\mathbf{S}^2/3$. 
Within this model of magnetic impurities the gapless superconductivity has been discussed \cite{Skalski1964a,Parks2018}.

The above models ignore the orientation anisotropy of the moments of magnetic impurities.
We argue that such an anisotropy is potentially relevant to the 2D superconductors that are of a current research interest \cite{Ugeda2015,Xi2015,Saito2016,Dvir2017,Liu2018,Sohn2018,Nakata2018,DelaBarrera2018}.
The magnetic impurities located at or near the interface between a 2D superconductor and adjacent monolayers are often anisotropic  due to the reduced spatial symmetry \cite{Vaz2008}.
If the spin of such an impurity, tend to point {\it e.g.} out of plane, the contribution of the out of plane spin-component to the scattering dominates the contribution of the in-plane components.

Recently, superconducting devices were fabricated by exfoliating one- to few- layers of transition metal dichalcogenides \cite{Geim2013}.
These systems often lack an inversion center.
This implies that the electronic bands are split by  spin-orbit coupling (SOC) which is an odd function of the electron momentum and hence, anisotropic.
Besides, the lack of an inversion symmetry leads to a parity mixed superconductivity with coexisting parity-even singlet and parity-odd triplet order parameters \cite{Gorkov2001,Yip2014a,Smidman2017}.
These 2D superconductors are essentially spin anisotropic and it is in this situation the out-of-plane and in-plane spin-components of impurities play a distinct role.

The superconductivity in 2D materials based on transition metal dichalcogenides with horizontal mirror symmetry, referred to as Ising superconductors withstand in-plane magnetic fields far beyond Pauli limit \cite{Bulaevskii976,Frigeri2006,Samokhin2008}.
Such a large field introduces a significant in-plane spin anisotropy. 
We demonstrate that this makes the two in-plane spin-components of impurities to be inequivalent in the way they affect the superconductivity.

To study the effect of the anisotropy of impurity spins we assume for clarity that different impurity spin-components are statistically independent.
In result, the total scattering rate due to magnetic impurities is $\Gamma_\mathrm{m}=\Gamma_1+\Gamma_2+\Gamma_3$, where  $\Gamma_i \propto \overline{S_i^2}$  originates from the spatial randomness of $i$-th spin-component of impurities.  
We study the effect of the magnetic impurity anisotropy by considering the scattering rates $\Gamma_i$ as independent  parameters. 
In materials where $\boldsymbol{\sigma}\cdot \mathbf{S}$ is the only spin-dependent interaction, the magnetic anisotropy of impurity spins is inconsequential.
In this case the total scattering rate $\Gamma_m$ characterizes the effectiveness of the magnetic impurities in suppressing the superconductivity, see Table~\ref{tab:intro}.
We demonstrate that in the presence of SOC and in-plane magnetic field makes all three spin-components of magnetic impurities distinct.

Recently, the authors demonstrated that the combined action of an in-plane magnetic field and Ising-type SOC converts isotropic singlet $s$-wave Cooper pairs to equal-spin triplet pairs \cite{PhysRevB.99.180505_mockli_2019,PhysRevB.101.014510_mockli2020}. It is then natural to ask, if time-reversal symmetry breaking by a magnetic field can cause singlet to triplet conversion, could magnetic impurities also cause conversion? 
We demonstrate that although each impurity contribute a local exchange field, on average the effect of such local field cancels out in the absence of a total net spin-polarization of magnetic impurities, see Sec. \ref{sec:conversion} for further details.

The paper is organized as follows. In Sec. \ref{sec:model}, we introduce our model Hamiltonian. In Sec. \ref{sec:Eilenberger}, we assume that the Fermi energy is the largest energy scale and develop the quasi-classical theory of the superconducting state. We obtain the Eilenberger equation that includes the effects of arbitrary spin-fields and impurities. Next, in Sec. \ref{sec:ising} we specialize to the case with Ising-SOC and an in-plane Zeeman field and obtain the corresponding Eilenberger equation that describes the superconducting transition. 
In Sec. \ref{sec:singlet} we address the case when there is pairing only in the singlet channel. We obtain the transition lines in the magnetic field--temperature phase diagram for both isotropic ($s$-wave, $d$-wave, etc...) and anisotropic ($d$-wave, $h$-wave, etc...) order-parameters. 
In Sec. \ref{sec:triplet} we address the case when there is pairing in the triplet channels only and discuss the role of the Cooper pair spin-polarization. 
In Sec. \ref{sec:crossovers}, we consider the case when the singlet $s$-wave pairing channel is dominant, but a sub-dominant triplet channel exists. Then, the increase of magnetic impurities can drive a crossover from a pure singlet to a pure triplet state. 
In Sec. \ref{sec:conversion} we analyze the case when singlet and triplet order-parameters coexist. The joint presence of SOC and Zeeman fields selects a specific triplet component to couple to the $s$-wave singlet. We discuss how the impurities affect this coupling. 
In Sec. \ref{sec:discussion}, we explain our results in the context of the current literature and conclude. 
In Appendix \ref{app:self_energy} we derive the impurity self-energy in the self-consistent Born approximation.

\begin{table}
\caption{\label{tab:intro}%
Summary of the effects of non-magnetic and magnetic impurities on isotropic (singlet) and anisotropic (singlet or triplet) superconducting states. $\alpha$ is the pair-breaking parameter that informs how strongly $T_\mathrm{c}$ is suppressed. The scalar (magnetic) scattering rate is denoted by $\Gamma_0$ ($\Gamma_\mathrm{m}$).}
\begin{ruledtabular}
\begin{tabular}{cccc}
Order-parameter & Condition & Non-magnetic & Magnetic  \\
\hline
Isotropic (singlet) & $\sum_\mathbf{k}\Delta(\mathbf{k})=\mathrm{const.}$ & $\alpha = 0$ & $\alpha = 2\Gamma_\mathrm{m}$  \\ 
Anisotropic & $\sum_\mathbf{k}\Delta(\mathbf{k})=0$& $\alpha =\Gamma_0+\Gamma_\mathrm{m}$ &  $\alpha =\Gamma_0+\Gamma_\mathrm{m}$  \\
\end{tabular}
\end{ruledtabular}
\end{table}

\section{The model}
\label{sec:model}

We employ a coordinate system such that the 2D superconductor lies in the $xy$-plane.
The Hamiltonian of a generic disordered non-centrosymmetric superconductor in the presence of a magnetic Zeeman field is \cite{Bauer2012}
\begin{align}
\mathcal{H} & = \sum_{\mathbf{k},s}\xi(\mathbf{k})c^\dag_{\mathbf{k}s}c_{\mathbf{k}s}
+\sum_{\mathbf{k},ss'}\left(\boldsymbol{\gamma}(\mathbf{k})-\mathbf{B} \right )\cdot\boldsymbol{\sigma}_{ss'}c^\dag_{\mathbf{k}s}c_{\mathbf{k}s'} \notag  \\
& +\frac{1}{2}\sum_{\mathbf{k},\mathbf{k}'}\sum_{\{s_i\}}V^{s_1s_2}_{s_1's_2'}\left(\mathbf{k},\mathbf{k}' \right )c^\dag_{\mathbf{k}s_1}c^\dag_{-\mathbf{k}s_2}c_{-\mathbf{k}'s_2'}c_{\mathbf{k}'s_1'} \notag \\
& +\frac{1}{2}\sum_{\mathbf{k},\mathbf{k}'}\sum_{ss'} U_{ss'}(\mathbf{k}-\mathbf{k}')c^\dag_{\mathbf{k}s}c_{\mathbf{k}'s'}.
\label{Hamiltonian}
\end{align}
Here, $\xi(\mathbf{k})=\xi(-\mathbf{k})$ is the symmetric part of the normal state dispersion counted from the chemical potential.
The vector of Pauli matrices is denoted by $\boldsymbol{\sigma}=(\sigma_1,\sigma_2,\sigma_3)$ with subscripts denoting the spatial directions $(x,y,z)$.
The anti-symmetric part $\boldsymbol{\gamma}(\mathbf{k})=-\boldsymbol{\gamma}(-\mathbf{k})$ is the spin-orbit coupling (SOC) that arises in crystals lacking an inversion center.
We indicate the Fermi surface average of the SOC as $\langle |\boldsymbol{\gamma}(\mathbf{k})|^2\rangle_\mathrm{FS}=\Delta^2$.
Here, the magnetic field $\mathbf{B}$ absorbs the prefactor $g\mu_\mathrm{B}/2$ with the $g$-factor and the Bohr magneton, and has dimension of energy. 
The Zeeman field breaks the time-reversal symmetry of the Hamiltonian. 
The superconducting pairing interaction $V^{s_1s_2}_{s_1's_2'}\left(\mathbf{k},\mathbf{k}' \right )$ includes all the pairing channels allowed by crystal symmetry.

We consider scalar and magnetic impurities distributed randomly and independently in the system.
The potential produced by a scalar and magnetic impurities are given by $U_0(\mathbf{k}-\mathbf{k}')\delta_{ss'}$ and $J(\mathbf{k}-\mathbf{k}')\mathbf{S}\cdot\boldsymbol{\sigma}_{ss'}$, respectively. 
We assume that the spins of magnetic impurities at different sites are uncorrelated.
Furthermore, we consider the distribution of impurities' spins and the distribution of their spatial location as statistically independent. 
This allows us to reduce the averaging over the magnetic impurities to averaging over the spin orientations and averaging over the spatial locations done independently.

\section{The Eilenberger equation}

\subsection{General theory\label{sec:Eilenberger}}

We use the Pauli-matrices $\{\rho_i\}$ (plus identity $\rho_0$) to generate Nambu-space, and $\{\sigma_i\}$ for spin-space. After a mean-field decoupling of the superconducting term in the Hamiltonian \eqref{Hamiltonian}, we write the Hamiltonian in $4\times 4$ matrix form in the basis defined by the Kronecker product $\rho_i\otimes \sigma_j$ ($\rho_i\sigma_j$ for short) \cite{Fulde1966a}. The matrix of the impurity part is \cite{Fulde1966a,Maekawa1987,Pint1989}
\begin{align}
\hat{U}(\mathbf{k}-\mathbf{k}')=U_0(\mathbf{k}-\mathbf{k}')\rho_3 \sigma_0+J(\mathbf{k}-\mathbf{k}')\mathbf{S}\cdot\boldsymbol{\alpha}, \label{Umatix}
\end{align}
where, for brevity, we have combined the contributions of the statistically independent scalar and magnetic impurity potentials, and $\boldsymbol{\alpha}=
\left(\rho_3 \sigma_1,\rho_0\sigma_2,\rho_3 \sigma_3 \right )$. 
For simplicity we consider a short range scattering impurity potential such that the scattering amplitudes in Eq.~\eqref{Umatix} are momentum independent, $U_0(\mathbf{k}-\mathbf{k}') = U_0$ and  $J(\mathbf{k}-\mathbf{k}') = J$.
In what follows, we keep track of the impurity spin components $\mathbf{S}=(S_1,S_2,S_3)$ explicitly, which will allow us to identify special effects associated to specific components $S_i$.

Assuming that the Fermi energy $E_\mathrm{F}$ is the largest energy scale,
we derive the quasi-classical theory for the Hamiltonian in Eq. \eqref{Hamiltonian}. 
We follow the same notations and definitions as in Ref. \cite{PhysRevB.101.014510_mockli2020}.
The central quantity in the quasi-classical theory is the normalized $4\times 4$ propagator that depends on the Fermi momentum $\mathbf{k}=\mathbf{k}_\mathrm{F}$ and the Matsubara frequencies $\omega_n=(2n+1)\pi T$
\begin{align}
    \hat{g}(\mathbf{k},\omega_n) =
\begin{bmatrix}
g(\mathbf{k},\omega_n) & -if(\mathbf{k},\omega_n)\\ 
-if^*(-\mathbf{k},\omega_n) & -g^*(-\mathbf{k},\omega_n)
\end{bmatrix},
\end{align}
where $g(\mathbf{k};\omega_n)=g_0(\mathbf{k};\omega_n)\sigma_0+\mathbf{g}(\mathbf{k};\omega_n)\cdot\boldsymbol{\sigma}$ and $f(\mathbf{k};\omega_n)=\left[f_0(\mathbf{k};\omega_n)\sigma_0+\mathbf{f}(\mathbf{k};\omega_n)\cdot\boldsymbol{\sigma})\right]i\sigma_2$. The propagator is found by solving the commutator Eilenberger equation
\begin{align}
\left[\left(i\omega_n\rho_0\sigma_0-\hat{\Sigma}(\omega_n)-\hat{S}(\mathbf{k})-\hat{V}(\mathbf{k}) \right )\rho_3\sigma_0,\hat{g}(\mathbf{k};\omega_n) \right ]=0,
\label{EilenbergerEq}
\end{align}
together with the normalization condition $\hat{g}^2(\mathbf{k};\omega_n)=\rho_0\sigma_0$.
The impurity self-energy $\hat{\Sigma}(\omega_n)$ is obtained within the self-consistent Born approximation (see Appendix \ref{app:self_energy} for details) and has the components $\hat{\Sigma}=\hat{\Sigma}_0+\hat{\Sigma}_1+\hat{\Sigma}_2+\hat{\Sigma}_3$ defined by
\begin{align}
 \hat{\Sigma}_0(\omega_n)& =-i\Gamma_0\rho_0\sigma_0\langle \hat{g}(\mathbf{k};\omega_n)\rangle_\mathrm{FS} \rho_3\sigma_0;\label{Sigma0} \\
 \hat{\Sigma}_1(\omega_n)& =-i\Gamma_1\rho_0\sigma_1\langle \hat{g}(\mathbf{k};\omega_n)\rangle_\mathrm{FS} \rho_3\sigma_1;\label{Sigma1} \\
  \hat{\Sigma}_2(\omega_n)& =-i\Gamma_2\rho_3\sigma_2\langle \hat{g}(\mathbf{k};\omega_n)\rangle_\mathrm{FS} \rho_0\sigma_2;\label{Sigma2} \\
  \hat{\Sigma}_3(\omega_n)& =-i\Gamma_3\rho_0\sigma_3\langle \hat{g}(\mathbf{k};\omega_n)\rangle_\mathrm{FS} \rho_3\sigma_3.\label{Sigma3}
\end{align}
The scalar impurity scattering rate is $\Gamma_0=\pi n_0 N_0U_0^2$, where $n_0$ is the number of scalar impurities, $N_0$ is the density of states per spin at the Fermi level.
The magnetic scattering rate components are $\Gamma_i = \pi n_\mathrm{m} N_0 J^2 \overline{ S_i^2 }$, where $n_\mathrm{m}$ is the number of magnetic impurities. 
For isotropic distribution of impurities' spins we have $\Gamma_1=\Gamma_2=\Gamma_3=\Gamma_\mathrm{m}/3$ with $\Gamma_\mathrm{m}=\pi n_\mathrm{m} N_0 J^2S(S+1)$, where $S$ is the total impurity spin. 
When the distribution of the spins of impurities is isotropic, we have 
$\Gamma_1=\Gamma_2=\Gamma_3=\Gamma_\mathrm{m}/3$ with $\Gamma_\mathrm{m}=\pi n_\mathrm{m} N_0 J^2S(S+1)$. In the more generic situation considered here, the scattering rates $\Gamma_i$ can be different.
The spin-fields enter Eq.~\eqref{EilenbergerEq} through the matrix
\begin{align}
\hat{S}(\mathbf{k})=
\begin{bmatrix}
(\boldsymbol{\gamma}(\mathbf{k})-\mathbf{B})\cdot\boldsymbol{\sigma} & 0_{2\times 2}\\ 
0_{2\times 2} & (\boldsymbol{\gamma}(\mathbf{k})+\mathbf{B})\cdot\boldsymbol{\sigma}^\mathrm{T},
\end{bmatrix}
\end{align}
and the superconducting order-parameters enters Eq.~\eqref{EilenbergerEq} through the matrix
\begin{align}
 \hat{V}(\mathbf{k})
=\begin{bmatrix}
0_{2\times 2} & \Delta(\mathbf{k}) \\ 
\Delta^\dag(\mathbf{k}) & 0_{2\times 2}
\end{bmatrix},
\end{align}
where $\Delta(\mathbf{k}) =\left[\psi(\mathbf{k})\sigma_0+\mathbf{d}(\mathbf{k})\cdot\boldsymbol{\sigma} \right ]i\sigma_2$ \cite{RevModPhys.63.239_sigrist,mineev1999introduction,Muzikar1999,doi:10.1146/annurev-conmatphys-031113-133912_yip}.
The Pauli principle enforces the function $\psi(\mathbf{k})$ parametrizing Cooper spin-singlets to be even, and the d-vector $\mathbf{d}(\mathbf{k})$ parametrizing Cooper spin-triplets to be odd. The diagonal elements of $\Delta(\mathbf{k})$ describes the $S=1$ triplets, and the off-diagonal elements contain the $S=0$ singlets and triplets. 
These order-parameters are related to the anomalous propagators $\{f_0,\mathbf{f}\}$ through the self-consistency condition given by \cite{PhysRevB.101.014510_mockli2020}
\begin{align}
&d_i(\mathbf{k}) \log\frac{T}{T_\mathrm{c}} 
\notag \\ & 
+\pi T\sum_{n=-\infty}^\infty\left[\frac{d_i(\mathbf{k}) }{|\omega_n|}-\hat{d}_i(\mathbf{k})\left\langle \hat{d}_i(\mathbf{k}') f_i(\mathbf{k}';\omega_n)\right\rangle_\mathrm{FS}  \right ]=0,
\label{selfconsistency}
\end{align}
where $\hat{d}_i(\mathbf{k})$ is a basis function of the corresponding order-parameter $d_i(\mathbf{k})$ that belongs to a specific irreducible representation of the crystal point group, and $T_\mathrm{c}$ is the superconducting transition temperature of the pairing channel. The same self-consistency condition holds for $\psi(\mathbf{k})$ with basis function $\hat{\psi}(\mathbf{k})$. 
In Eq.~\eqref{selfconsistency} the basis functions are normalized, $\langle \hat{d}^2_i(\mathbf{k})\rangle_{\mathrm{FS}} = \langle \hat{\psi}^2(\mathbf{k})\rangle_{\mathrm{FS}} = 1$, where $\langle \ldots \rangle_{\mathrm{FS}}$ stands for the angular averaging over the directions of $\hat{\mathbf{k}}$, $\int_0^{2\pi}\frac{\mathrm{d}\varphi_\mathbf{k}}{2\pi} (\ldots) $.

\begin{figure}
\centering
\includegraphics[width=0.45\textwidth]{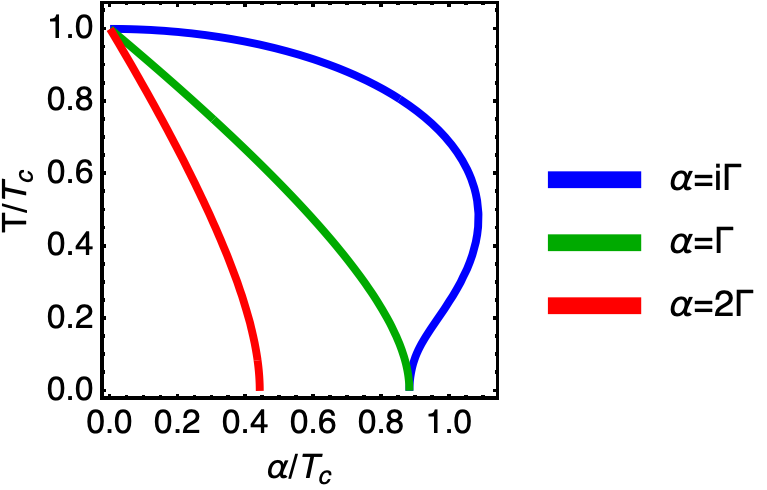}
\caption{\label{fig:0} 
Contour plot of the pair-breaking equation \eqref{AG} showing the suppression of the superconducting critical temperature $T/T_\mathrm{c}$ for different pair-breaking parameters $\alpha$. The blue curve shows the case for a purely imaginary pair-breaking parameter, which is typically the case for spin-fields (magnetic field and SOC). The green and the red curves show the case for a purely real parameter, which is typically the case for impurities.
}
\end{figure}

To obtain the pair-breaking equation that describes the superconducting transition $T(B)$, we solve the linearized Eilenberger equation for the superconducting propagators $\{f_0,\mathbf{f}\}$, substitute the solutions into the self-consistency conditions \eqref{selfconsistency}, and perform the summation over the Matsubara frequencies.
The result of this summation
contains all the information on how the superconducting order-parameters are affected by the spin-fields and impurities.
Eq. \eqref{selfconsistency} can then be written in the form $\log(T/T_\mathrm{c})+\mathcal{S}=0$, where $\mathcal{S}$ is a Matsubara sum.
In simple special cases, such as when the order parameter is purely singlet/triplet and in the absence of one of the spin-fields, the Matsubara sum can be performed analytically and the pair-breaking equation adopts the form \cite{tinkham2004introduction}
\begin{align}
\log\frac{T}{T_\mathrm{c}}+\mathrm{Re}\,\psi\left(\frac{1}{2}+\frac{\alpha}{2\pi T} \right )-\psi \left(\frac{1}{2} \right ) = 0,
\label{AG}
\end{align}
where $\alpha$ is a generic pair-breaking parameter that can be a combination of scattering rates and spin-fields, $\psi(z)$ is the digamma function and $T$ is the new $\alpha$-affected critical temperature. We plot the contour of Eq. \eqref{AG} in Fig. \ref{fig:0}.
In the absence of spin-fields, only magnetic impurities are pair-breakers in isotropic singlet superconductors with $\alpha=2\Gamma_\mathrm{m}$. 
In anisotropic singlet superconductors, both scalar and magnetic impurities are equally pair-breaking with $\alpha=\Gamma_0+\Gamma_\mathrm{m}$. Since isotropic singlets suffer twice as much from magnetic impurities than anisotropic singlets, the effect of magnetic impurities can give indications of isotropy/anisotropy of the superconducting state. In triplet superconductors $\alpha=\Gamma_0+\Gamma_\mathrm{m}$, which is the same as in anisotropic singlet superconductors. This changes in the presence of SOC and Zeeman field.

\subsection{The case of Ising superconductivity \label{sec:ising}}

Up to now, the discussion applies generically to non-centrosymmetric superconductors with an arbitrary direction of the spin-fields. A case of special interest is the one when the magnetic field is in-plane, such that orbital depairing effects can be neglected. 
Moreover, if the SOC is out-of-plane, its action on the Cooper pair spins makes them more robust against paramagnetic depairing by an in-plane magnetic field. A similar situation with $\boldsymbol{\gamma}(\mathbf{k})\perp \mathbf{B}$ in a 2D superconductor can also be realized when $\boldsymbol{\gamma}(\mathbf{k})$ is restricted to the plane, such as helical Rashba SOC, and $\mathbf{B}$ applied perpendicularly to the plane. In this case, the orbital depairing mechanism is expected to play a dominant role. In what follows, we specialize to the Ising-SOC with an in-plane Zeeman field. In fact, an Ising superconductor is the simplest non-centrosymmetric superconductor because it only has one SOC component $(0,0,\gamma_3(\mathbf{k}))$ as opposed to Rashba (or more complicated SOCs) which has two components $(\gamma_1(\mathbf{k}),\gamma_2(\mathbf{k}),0)$. Therefore, without loss of generality for the Ising case, we henceforth set $\mathbf{B}=(B,0,0)$ and $\boldsymbol{\gamma}(\mathbf{k})=(0,0,\gamma(\mathbf{k}))$.

To study the superconducting transition, we linearize the Eilenberger equation \eqref{EilenbergerEq} and write it in matrix form as 
\begin{align}
\label{Eilenberger}
\begin{bmatrix}
\tilde{\omega}_n & -iB & 0 & 0\\ 
-iB & \tilde{\omega}_n & \gamma(\mathbf{k}) & 0\\ 
0 & -\gamma(\mathbf{k}) & \tilde{\omega}_n & 0\\ 
0 & 0 & 0 & \tilde{\omega}_n
\end{bmatrix}
\begin{bmatrix}
f_0(\mathbf{k};\omega_n)\\ 
f_1(\mathbf{k};\omega_n)\\ 
f_2(\mathbf{k};\omega_n)\\ 
f_3(\mathbf{k};\omega_n)
\end{bmatrix}
=s
\begin{bmatrix}
\tilde{\psi}(\mathbf{k};\omega_n)\\ 
\tilde{d}_1(\mathbf{k};\omega_n)\\ 
\tilde{d}_2(\mathbf{k};\omega_n)\\ 
\tilde{d}_3(\mathbf{k};\omega_n)
\end{bmatrix},
\end{align}
where $s=\mathrm{sgn}(\omega_n)$ and the impurity rescaled quantities are defined by
\begin{align}
& \tilde{\omega}_n  =\omega_n+s\left(\Gamma_0+\Gamma_\mathrm{m}\right ); \\
& \tilde{\psi}(\mathbf{k};\omega_n)  = \psi(\mathbf{k})+\left(\Gamma_0-\Gamma_\mathrm{m} \right )\langle f_0(\mathbf{k};\omega_n)\rangle_\mathrm{FS}; \\
& \tilde{d}_i(\mathbf{k};\omega_n) = d_i(\mathbf{k})+\Gamma_i'\langle f_i(\mathbf{k};\omega_n)\rangle_\mathrm{FS},
\end{align}
with
\begin{align}
& \Gamma_0'=\Gamma_0-\Gamma_1-\Gamma_2-\Gamma_3 ; \\
& \Gamma_1' = \Gamma_0-\Gamma_1+\Gamma_2+\Gamma_3 ; \\
&  \Gamma_2' = \Gamma_0+\Gamma_1-\Gamma_2+\Gamma_3 ; \\
& \Gamma_3' = \Gamma_0+\Gamma_1+\Gamma_2-\Gamma_3. 
\end{align}
Here, $\Gamma_\mathrm{m} = \Gamma_1+\Gamma_2+\Gamma_3$ while we are not making an assumption of the components $\Gamma_i$ being equal. 
We also introduce the notation $\Gamma = \Gamma_0+\Gamma_\mathrm{m}$.

In the following sections we present the results for solving special cases of our master equation \eqref{Eilenberger}. We consider the purely singlet, purely triplet, and the general singlet-triplet mixed cases. We also discuss the possibility of magnetic impurity induced crossovers from leading isotropic singlet states to sub-leading anisotropic states.

\begin{figure*}
\centering
\includegraphics[width=0.8\textwidth]{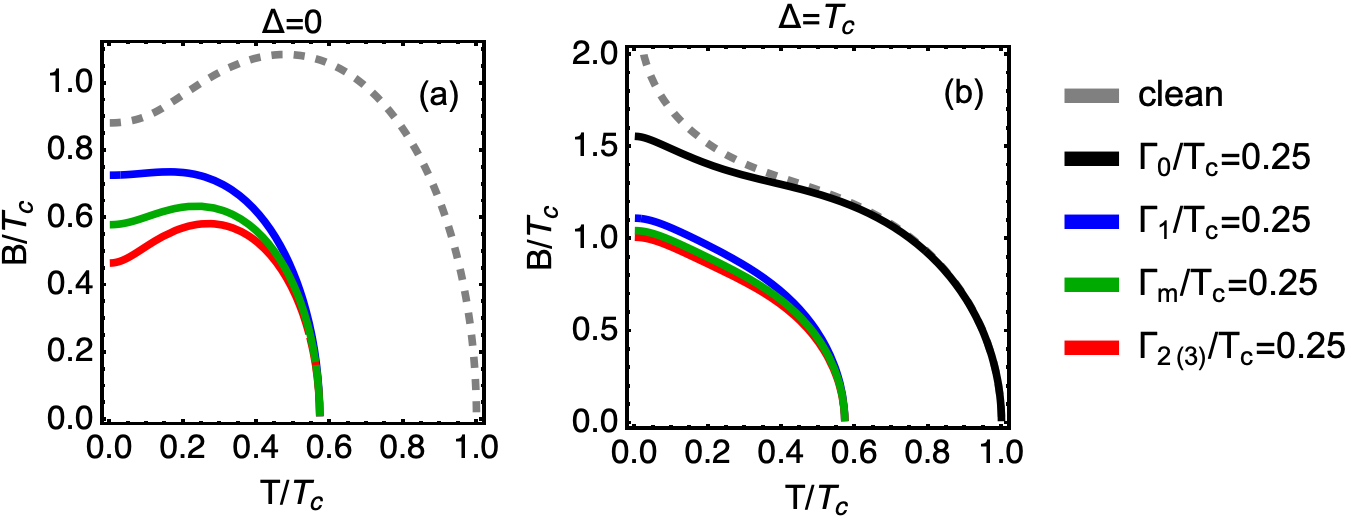}
\caption{\label{fig:1} 
The effect of magnetic and non-magnetic impurities on purely isotropic ($s$-wave) singlet superconductors. The magnetic impurity component $\Gamma_1$ that is parallel to the magnetic field has a weaker effect than the perpendicular components $\Gamma_{2(3)}$. The different effects of $\Gamma_i$ ($i=1,2,3$) become less relevant for larger values of SOC. For the green curved we used $\Gamma_1=\Gamma_2=\Gamma_3=\Gamma_\mathrm{m}/3$.
}
\end{figure*}

\section{Singlet superconductors \label{sec:singlet}}

We examine two cases
for the momentum structure $\psi(\mathbf{k})$ of the singlet order-parameters. Case (A): the order-parameter is isotropic ($s$-wave) $\psi(\mathbf{k})=\psi_0$, where $\psi_0$ is a constant (the basis function $\hat{\psi}(\mathbf{k})=1$). 
Case (B): the order-parameter is anisotropic with $\langle \psi(\mathbf{k})\rangle_\mathrm{FS}=0$ and the basis function $\hat{\psi}(\mathbf{k})$ is even. Below we analyze these cases.

\subsection{Isotropic order parameter}

We set $d_i(\mathbf{k})=0$ in Eq. \eqref{Eilenberger} and using $\gamma(\mathbf{k})=\Delta\mathrm{sgn}[\gamma(\mathbf{k})]$ for simplicity, the pair-breaking equation for $\psi_0$ reads $\log(T/T_\mathrm{c})+\mathcal{S}_\mathrm{s}=0$ with 
\begin{align}
\mathcal{S}_\mathrm{s} & = 
\pi T\sum_{n=-\infty}^\infty
\biggr[
\frac{1}{|\omega_n|} \label{Ss} \\
&
-\frac{|\tilde{\omega}_n|(|\omega_n|+2\Gamma_1)+\Delta^2 }{|\tilde{\omega}_n|B^2+(|\omega_n|+2\Gamma_m)\left[|\tilde{\omega}_n|(|\omega_n|+2\Gamma_1)+\Delta^2 \right ]} 
\biggr]. \notag 
\end{align}
In the absence of the Zeeman field $B$, the pair-breaking equation reduces to Eq. \eqref{AG} with $\alpha=2\Gamma_\mathrm{m}$; see the red curve in Fig. \ref{fig:0}. In this case,
all magnetic components of $\Gamma_i$ ($i=1,2,3$) are equally detrimental to superconductivity.
In the low temperature limit, superconductivity is obliterated for $\alpha/T_\mathrm{c}=\pi/(2e^\gamma)\approx 0.88$, where $\gamma\approx 0.58$ is the Euler constant. The constant $\gamma$ always appears in the exponent and should not cause any confusion with SOC.

\subsubsection{Inversion symmetric case, \texorpdfstring{$\Delta=0$}{}}

In Fig. \ref{fig:1}a we show the case with inversion-symmetry, for which $\Delta=0$. 
The clean (gray) curve is described by the imaginary pair-breaking parameter $\alpha=iB$.
By setting $\Delta=0$ in Eq. \eqref{Ss}, one can see that superconductivity remains indifferent to scalar impurities. All the magnetic components act as pair-breakers, but in different ways. 
The magnetic impurities $\Gamma_1$ that are parallel to the Zeeman field have a weaker effect than the magnetic impurities with perpendicular components $\Gamma_{2(3)}$. To show this we first plot the case when the magnetic impurity directions are randomly oriented, such that $\Gamma_1=\Gamma_2=\Gamma_3=\Gamma_\mathrm{m}/3$; see the green curve in Fig. \ref{fig:1}a.
If we now consider anisotropic magnetic impurities with all impurity spins $\Gamma_1$ parallel to the Zeeman field $B$, the critical field at low temperatures is higher; see the blue curve. Similarly, for spin impurities $\Gamma_{2(3)}$ aligned only perpendicularly to the Zeeman field, suppression is maximized; see the red curve. 
These perpendicular spin impurities add a new direction of depairing, making them more detrimental.

\subsubsection{The case of no inversion symmetry, \texorpdfstring{$\Delta \neq 0$}{}}

In Fig. \ref{fig:1}b we show the case without inversion-symmetry by adding $\Delta=T_\mathrm{c}$. The SOC enhances the critical field because it counteracts the Zeeman field. 
In contrast to the case with inversion, non-magnetic impurities $\Gamma_0$ now suppress the critical field (black curve), because they undo the enhancement caused by SOC \cite{PhysRevB.101.014510_mockli2020}.
The larger the SOC, the harder it becomes to distinguish the different effects of the spin impurity components $\Gamma_i$.

\subsection{Anisotropic order parameter}

\begin{figure}
\centering
\includegraphics[width=0.48\textwidth]{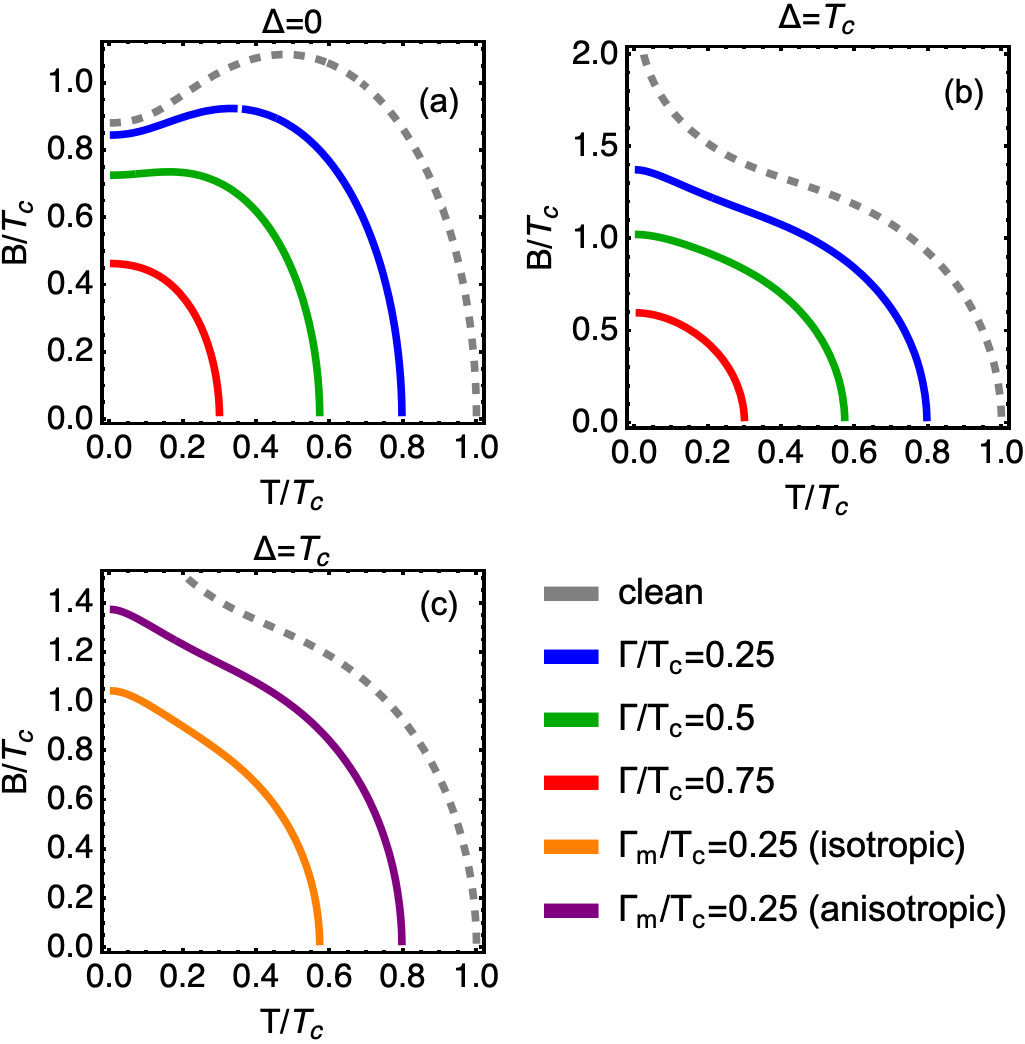}
\caption{\label{fig:2} 
The effect of impurities on purely anisotropic singlet ($d$-wave, $h$-wave, etc...) superconductors. Magnetic and non-magnetic impurities have the same effect, which is indicated by a generic scattering rate $\Gamma$. In (c) we compare the effect of a magnetic scattering rate $\Gamma_\mathrm{m}$ on an isotropic singlet (orange) and anisotropic singlet (purple) state assuming that both of their clean $T_\mathrm{c}$ is the same. 
}
\end{figure}

The Matsubara sum for the case with $\psi(\mathbf{k})$ is

\begin{align}
\mathcal{S}_\mathrm{s}= 
\pi T\sum_{n=-\infty}^\infty
\left[
\frac{1}{|\omega_n|} 
-\frac{\tilde{\omega}_n^2+\Delta^2}{|\tilde{\omega}_n|(\tilde{\omega}_n^2+B^2+\Delta^2)}
\right].
\label{SsAnisotripic}
\end{align}
The sum can be performed exactly, but we choose to maintain it in this simple form for discussion purposes. In contrast to the case of isotropic singlet order parameter, now any type of impurity is equally detrimental to superconductivity. 
In the absence of the Zeeman field, the pair-breaking equation reduces to Eq. \eqref{AG} with $\alpha=\Gamma_0+\Gamma_\mathrm{m}$. 
Anisotropic order-parameters have an intrinsic phase structure, which makes the dephasing by magnetic impurities less effective as compared to isotropic order-parameters. 
In Fig. \ref{fig:2}(a,b) we show the effect of arbitrary impurities $\Gamma$ on the transition line with inversion symmetry ($\Delta=0$) and without inversion symmetry ($\Delta=T_\mathrm{c}$), respectively. 
Throughout the paper, we use $\Gamma$ without a subscript if the nature of the impurities is unimportant.
In Fig. \ref{fig:2}c we compare the transition lines of an isotropic singlet order-parameter $\psi_0$ (orange) to an anisotropic order-parameter $\psi(\mathbf{k})$ for an equal amount of magnetic impurities $\Gamma_\mathrm{m}$. Isotropic superconductivity is obliterated for $\Gamma_\mathrm{m}/T_\mathrm{c}\approx 0.44$ and anisotropic for $\Gamma_\mathrm{m}/T_\mathrm{c}\approx 0.88$.

\section{Triplet superconductors \label{sec:triplet}}

In triplet superconductors, the order-parameter is the three component vector $\mathbf{d}(\mathbf{k})=(d_1(\mathbf{k}),d_2(\mathbf{k}),d_3(\mathbf{k}))$. 
We set $\psi(\mathbf{k})=0$ in the linearized Eilenberger equation \eqref{Eilenberger}, which together with the self-consistency condition \eqref{selfconsistency} yields the pair-breaking equations for the d-vector components $d_i(\mathbf{k})$ ($i=1,2,3$), which read $\log(T/T_\mathrm{c})+\mathcal{S}_{\mathrm{t}i}=0$, with the sums given by
\begin{align}
\label{t1}
\mathcal{S}_{\mathrm{t}1} & = 
\pi T\sum_{n=-\infty}^\infty
\biggr[
\frac{1}{|\omega_n|} \\
& -\frac{|\tilde{\omega}_n|(|\omega_n|+2\Gamma_2)}{(|\omega_n|+2\Gamma_2)B^2+|\tilde{\omega}_n|\left[|\tilde{\omega}_n|(|\omega_n|+2\Gamma_2)+\Delta^2 \right ]}
\biggr]; \notag 
\end{align}
\begin{align}
\label{t2}
\mathcal{S}_{\mathrm{t}2} & = 
\pi T\sum_{n=-\infty}^\infty
\biggr[
\frac{1}{|\omega_n|} \\
& -\frac{(|\omega_n|+2\Gamma_1)(|\omega_n|+2\Gamma_\mathrm{m})+B^2}{|\tilde{\omega}_n|B^2+(|\omega_n|+2\Gamma_\mathrm{m})\left[|\tilde{\omega}_n|(|\omega_n|+2\Gamma_1)+\Delta^2 \right ]}
\biggr]; \notag 
\end{align}
\begin{align}
\label{t3}
\mathcal{S}_{\mathrm{t}3}= 
\pi T\sum_{n=-\infty}^\infty
\left[
\frac{1}{|\omega_n|} 
-\frac{1}{|\tilde{\omega}_n|}
\right],\quad (\alpha=\Gamma_0+\Gamma_\mathrm{m}).
\end{align}
For simplicity, we assume that the $T_\mathrm{c}$'s for all triplet components are the same.

\subsection{Analysis of the results}

Let us analyze these results in detail. First, note that in the absence of spin-fields ($B=\Delta=0$), the d-vector components are indistinguishable ($\mathcal{S}_{\mathrm{t}1}=\mathcal{S}_{\mathrm{t}2}=\mathcal{S}_{\mathrm{t}3}$), and the pair-breaking parameter is $\alpha=\Gamma_0+\Gamma_\mathrm{m}$, which is the same for the anisotropic singlet case. 
For finite spin-fields, the $d_3$ component is parallel to SOC and perpendicular to the Zeeman field,
such that it remains unaffected by them. Only impurities suppress the $d_3$ component. It is then reasonable to expect that in the presence of spin-fields the $d_3$ may become a dominate superconducting channel. 

\begin{figure}
\centering
\includegraphics[width=0.48\textwidth]{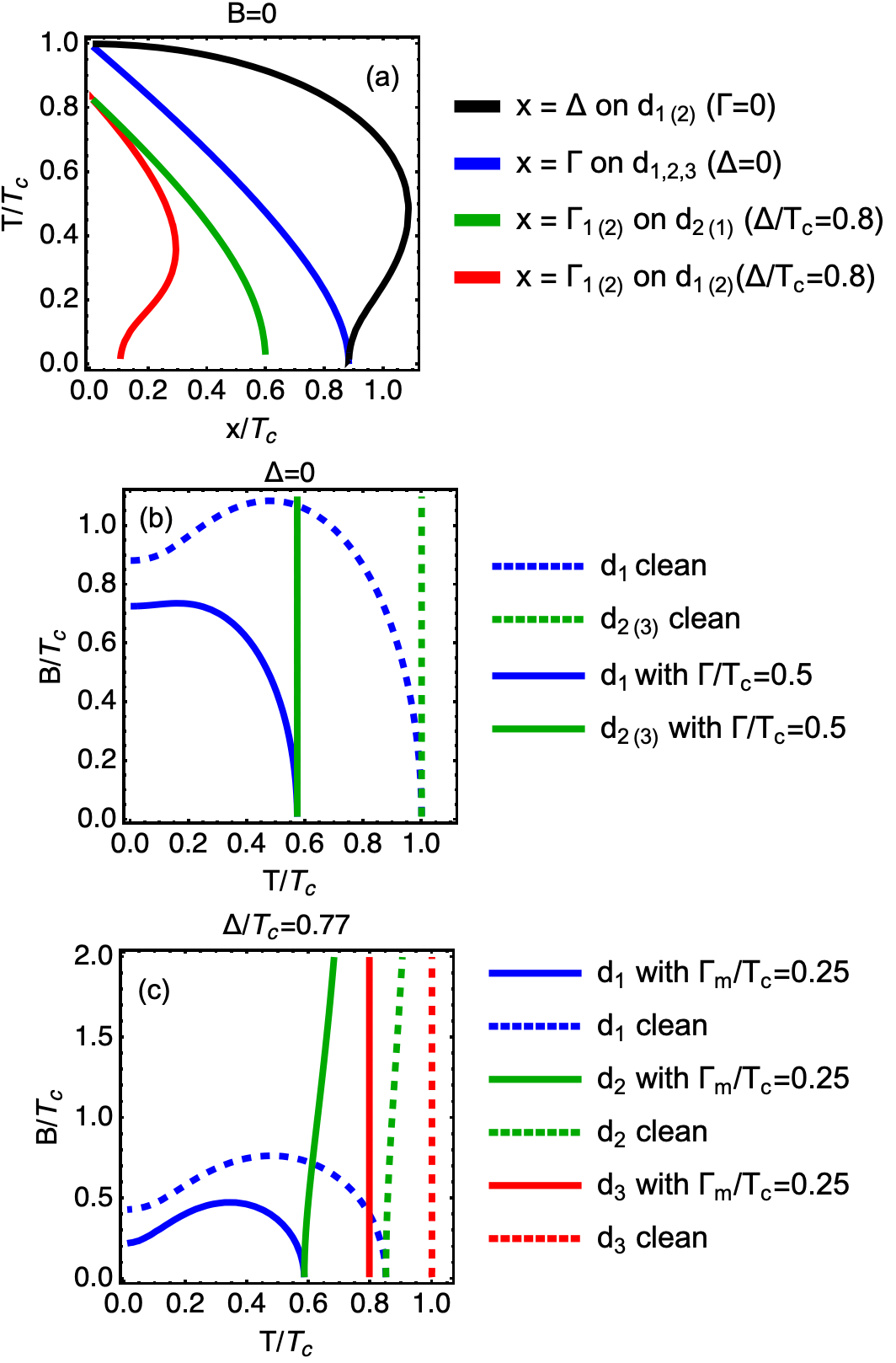}
\caption{\label{fig:3} 
The effect of impurities on purely triplet ($p$-wave, $f$-wave, etc...) superconductors. (a) The suppression of the critical temperature $T/T_\mathrm{c}$ as a function of $x$, where $x$ is a combination of an impurity scattering rate with SOC. The black curve shows how SOC is detrimental to the in-plane d-vector components and is described by $\alpha=i\Delta$. The blue curve shows how any impurities suppress the d-vector components in the absence of spin-fields. The green and red curves show how the in-plane $\Gamma_i$ affects the in-plane d-vectors components. Suppression is maximum (red curve) when $\Gamma_i$ is perpendicular to the Cooper pair polarization, and weaker (green) when $\Gamma_i$ is parallel to the pair polarization. 
(b) The effect of impurities in the presence of the magnetic field without SOC. The components $d_{2(3)}$ that are perpendicular to the Zeeman field remain unaffected by it and only suffer from impurities. The $d_1$ component that is parallel to the field suffers from both paramagnetic limiting and the impurities. 
(c) The effect of magnetic impurities when both the SOC and Zeeman fields are present. Here the value of SOC is fixed to $\Delta/T_\mathrm{c}=0.77$, such that the components $d_{1(2)}$ that are perpendicular to SOC are suppressed by SOC even at $B=0$ (see (a)). The order of increasing robustness to the impurities in the purely triplet case is: $d_1,d_2,d_3$. In particular, $d_2$ (green) displays a peculiar behavior at high magnetic fields. The green curves asymptotically approach the red curves for high magnetic fields. The Zeeman field undoes the suppression caused by SOC.
}
\end{figure}

To discuss how the spin-fields affect the in-plane d-vector components, let us consider the effect of SOC and the Zeeman field separately.
In Fig. \ref{fig:3}a we show the situation with time-reversal symmetry ($B=0$), but finite SOC. Eqs. \eqref{t1} and \eqref{t2} assume the form
\begin{align}
\label{eq:bzero}
\mathcal{S}_{\mathrm{t}1(2)}^{B\rightarrow0} = 
\pi T\sum_{n=-\infty}^\infty
\left[
\frac{1}{|\omega_n|} 
-\frac{|\omega_n|+2\Gamma_{2(1)}}{|\tilde{\omega}_n|(|\omega_n|+2\Gamma_{2(1)})+\Delta^2}
\right].
\end{align}
The presence of SOC gives a preferred spin-structure to Cooper pairs and makes the $d_i$ components inequivalent in how they respond to the impurity components $\Gamma_i$. In the clean case, $d_{1(2)}$ is suppressed by SOC with $\alpha=i\Delta$; see black curve in Fig.\ref{fig:3}a. 
With magnetic impurities, the $\Gamma_{1(2)}$ component has a weaker effect on $d_{2(1)}$ than $\Gamma_{1(2)}$ on $d_{1(2)}$; compare red and green curves in Fig. \ref{fig:3}a.

In Fig. \ref{fig:3}b we consider only the Zeeman field ($\Delta=0$), and we obtain the pair-breaking parameters $\alpha=\{iB+\Gamma_0+\Gamma_\mathrm{m},\Gamma_0+\Gamma_\mathrm{m},\Gamma_0+\Gamma_\mathrm{m}\}$ for $\{d_1,d_2,d_3\}$, respectively. Whereas the Zeeman field suppresses the parallel $d_1$ component, it has no effect on the perpendicular  components $d_{2,3}$.

We show the case with both SOC and Zeeman field in Fig. \ref{fig:3}c.
The $d_1$ component that points along the Zeeman field is the only one suppressed by it. It is also suppressed by SOC, such that even in the clean case the transition temperature for $d_1$ channel $T/T_\mathrm{c}<1$. We draw special attention to the behavior of $d_2$. 
The joint presence of SOC and the Zeeman field favors the $d_2$ component. 
An increasing magnetic field minimizes the suppression caused by SOC; see how the green curves asymptotically approach the vertical red lines in Fig. \ref{fig:3}c.

\subsection{Polarization of the triplets\label{sec:pol}}

A more intuitive understanding can be gained by interpreting the effects of the impurity components in terms of the Cooper pair spin polarization.
The momentum dependent spin-polarization of a Cooper pair is defined as the expectation value \cite{RevModPhys.63.239_sigrist}
\begin{align}
\mathbf{P}(\mathbf{k})= & \frac{1}{2}\mathrm{tr}\left[\Delta^\dag(\mathbf{k})\boldsymbol{\sigma}\Delta(\mathbf{k}) \right ]\notag \\
= &\psi(\mathbf{k})\mathbf{d}^*(\mathbf{k})+\psi^*(\mathbf{k})\mathbf{d}(\mathbf{k}) 
 + i\mathbf{d}(\mathbf{k})\times\mathbf{d}^*(\mathbf{k}).
 \label{polarization}
\end{align}
Since  $\psi(\mathbf{k})$ is even and $\mathbf{d}(\mathbf{k})$ is odd, only $\mathbf{q}=i\mathbf{d}(\mathbf{k})\times\mathbf{d}^*(\mathbf{k})$ potentially contributes to the total polarization averaged over the Fermi surface $\langle \mathbf{P}(\mathbf{k})\rangle_\mathrm{FS}$.
For the present pure triplet situation, $\psi(\mathbf{k})=0$. 
We now provide a heuristic motivation for the phases of $(d_1,d_2,d_3)$ to derive the polarization $\mathbf{P}=(P_1,P_2,P_3)$, which can also be obtained more rigorously \cite{PhysRevB.99.180505_mockli_2019}. Since SOC respects time-reversal, and the Zeeman field is in-plane, it is reasonable to expect $P_3=0$. This imposes $d_1d_2^*=d_2d_1^*$, such that both $d_1$ and $d_2$ are either purely real or purely imaginary. We know that $d_2$ is promoted by $B$, see numerator in \eqref{t2}, and is expected to break time-reversal. Therefore, we choose $d_{1(2)}(\mathbf{k})=i\eta_{1(2)}\hat{d}(\mathbf{k})$ to be imaginary, where $\eta_{1(2)}$ are real coefficients and $\hat{d}(\mathbf{k})$ are basis functions as in Eq. \eqref{selfconsistency}. The $d_3$ components can be chosen to be real, such that $d_3(\mathbf{k})=\eta_3\hat{d}(\mathbf{k})$. With this, we obtain the total polarization
\begin{align}
\label{eq:polarization}
    \langle \mathbf{P}(\mathbf{k})\rangle_\mathrm{FS} = 2\eta_3(-\eta_2,\eta_1,0),
\end{align}
where we used $\langle \hat{d}^2(\mathbf{k})\rangle_\mathrm{FS} = 1$. 
This tells us that $d_{1(2)}$ is responsible for a Cooper pair spin polarization along $y$ ($x$).

Now we present an intuitive picture of these results. 
We argue that in a in triplet superconductor, a finite Cooper pair spin polarization can explain why different spin-components of impurities have a distinct effect on superconductivity.
Recall that in the case of isotropic singlet order parameter, Eq. \eqref{Ss}, the impurity spin-component parallel to the applied magnetic field has a weaker effect on the critical field, compared to the perpendicular components. 
In a similar way, in the triplet superconductor with the Cooper pairs with a net polarization along a specific direction suffer less from the magnetic impurity component $\Gamma_i$ that is parallel to that direction. From Eq. \eqref{eq:polarization}, we see that $d_2$ is responsible for a Cooper pair polarization along $x$, such that, according to Eq. \eqref{eq:bzero}, the $\Gamma_{1(2)}$ component has a weaker (stronger) effect. Similarly, $\Gamma_{1(2)}$ has a stronger (weaker) effect on $d_1$ that is responsible for polarization along $y$. This can be summarized as follows: the $\Gamma_i$ component that is parallel to a net-field (either magnetic field or Cooper pair polarization) has a weaker effect.

\begin{figure*}
\centering
\includegraphics[width=0.8\textwidth]{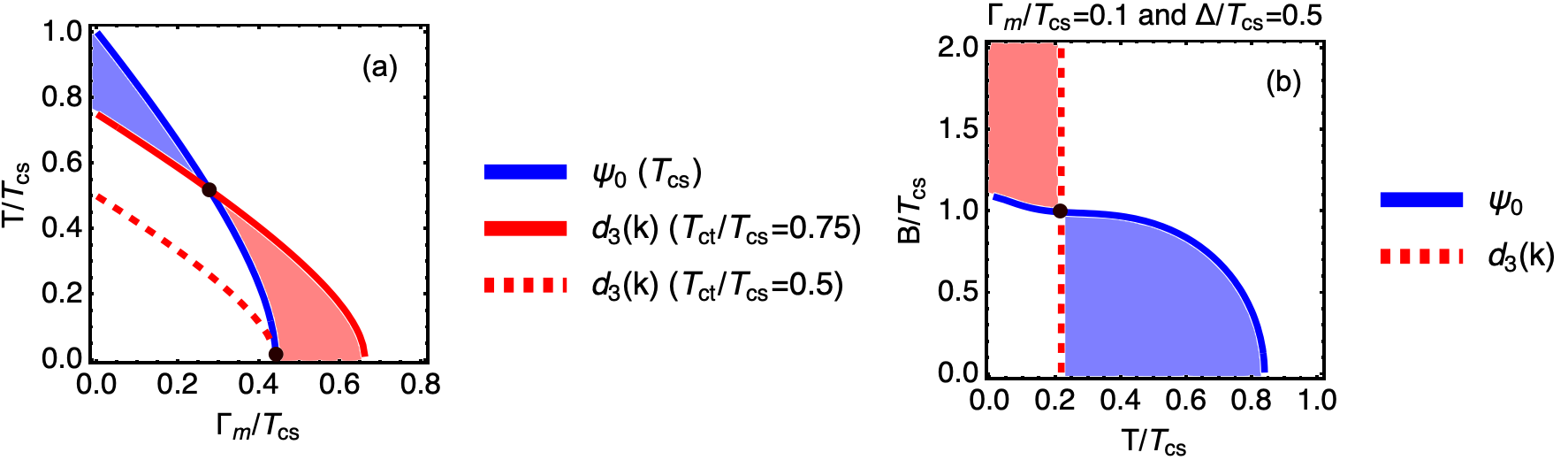}
\caption{\label{fig:4} 
Magnetic impurity induced crossover from a leading isotropic singlet $\psi_0$ ($s$-wave) to a sub-leading anisotropic triplet state $d_3(\mathbf{k})$ ($p$-wave, $f$-wave, etc...). (a) Possible crossovers at zero magnetic field.  
At zero magnetic field, a crossover happens for $T_\mathrm{ct}/T_\mathrm{cs}>1/2$ (see red-dashed curve). (b) At finite magnetic field, the $d_3$ triplets are always favorable below $T_\mathrm{ct}^*$, for field above the blue curve. 
The value $\Delta/T_\mathrm{cs}=0.5$ is only used for illustration purposes.
}
\end{figure*}

\section{Crossover from singlet to triplet superconductivity by magnetic impurities\label{sec:crossovers}}

In this section, we consider a leading isotropic singlet ($s$-wave) channel with superconducting critical temperature $T_\mathrm{cs}$, and a sub-leading triplet channel with corresponding critical temperature $T_\mathrm{ct}<T_\mathrm{cs}$. The discussion applies to both the cases with and without inversion.
For the arguments of this section, the effects of specific $\Gamma_i$ components is less relevant, such that we set $\Gamma_1=\Gamma_2=\Gamma_3=\Gamma_\mathrm{m}/3$.
Let us say that both singlet and triplet channels belong to the same crystal symmetry irreducible representation, such as $\psi_0$ and $d_3$ of the previous sections.
We also learned from the previous section that $d_3$ dominates over $d_1$ and $d_2$ in the presence of spin-fields, so it is reasonable to concentrate on $d_3$.

The singlet transition temperature $T_\mathrm{cs}$ suffers more from magnetic impurities $\Gamma_\mathrm{m}$ than the triplet transition temperature $T_\mathrm{ct}$, see Fig. \ref{fig:0}. With $T_\mathrm{cs}>T_\mathrm{ct}$, it is then possible to observe a magnetic impurity induced crossover from a singlet to a triplet state. 
The precise value of $T_\mathrm{ct}/T_\mathrm{cs}$ is an issue of pairing mechanism physics. Sub-leading triplet instabilities are present even in conventional pairing mechanisms \cite{Rosenstein2015}.
It is then natural to ask the guiding question of this section: what is the minimum ratio of $T_\mathrm{ct}/T_\mathrm{cs}$ to observe such a singlet to triplet crossover?

\subsection{No magnetic field}

At $B=0$, a crossover occurs if $T_\mathrm{ct}/T_\mathrm{cs}>1/2$. 
This result can be analytically obtained by combining two pair-breaking equations of the form in Eq. \eqref{AG}, one for the singlets and another for the triplets. This is illustrated in Fig. \ref{fig:4}a.
In the clean case,
the singlet (blue) curve starts from a critical temperature $T/T_\mathrm{cs}=1$, and is obliterated for $\Gamma_\mathrm{m}/T_\mathrm{cs}\approx 0.44$. The triplet (red) curve starts from $T/T_\mathrm{cs}=0.75$ and is obliterated for $\Gamma_\mathrm{m}/T_\mathrm{cs}\approx 0.66$. The blue region shows a purely singlet $\psi_0$ state and the red region shows a purely triplet $d_3(\mathbf{k})$ state. 
The precise crossover boundaries inside the phase diagram would depend on a treatment beyond linearization. 
The red-dashed curve shows the case with $T_\mathrm{ct}/T_\mathrm{cs}=1/2$.

\subsection{With magnetic field}

In Fig. \ref{fig:4}b we show the case at finite magnetic fields for a fixed value of magnetic impurities $\Gamma_\mathrm{m}/T_\mathrm{cs}=0.1$. The value of the ratio is set to $T_\mathrm{ct}/T_\mathrm{cs}=0.3$, such that no crossover at zero field is possible. 
Because of the magnetic impurities, the effective singlet (triplet) transition temperatures $T^*_\mathrm{cs}$ ($T^*_\mathrm{ct}$) start at lower values. Since the $d_3$ triplets remain unaffected by spin-fields, it will always be the leading instability at high magnetic fields below $T^*_\mathrm{ct}$.

\section{Singlet-triplet conversion by spin-fields \label{sec:conversion}}

In the previous sections, the superconducting order-parameter was either a pure singlet or a pure triplet. In this section, we address the the general situation with coexisting order-parameters $\{\psi(\mathbf{k}),\mathbf{d}(\mathbf{k})\}$. To do this, we study
the general solution of the Eilenberger equation \eqref{Eilenberger}.

A coupling between singlet and triplet order-parameters can originate from mainly two reasons:
(i) The densities of states of the spin-split bands are different \cite{Frigeri2004,Frigeri2006,Smidman2017}. This would lead to a coupling of $\psi_0$ and $d_3(\mathbf{k})$, which belongs to the same irreducible representation. Here, we neglect the possible difference of the densities of states, such that $d_3(\mathbf{k})$ remains decoupled. This information is already built into the Eilenberger equation \eqref{Eilenberger};
(ii) The joint action of SOC and magnetic field component that are perpendicular to each other. This can be seen from the matrix structure of the linearized Eilenberger equation \eqref{Eilenberger}. The presence of both spin-fields will select the $d_2(\mathbf{k})$ triplets to couple to the $\psi_0$ singlets. We have pointed this out in our previous works, and we refer to Refs. \cite{PhysRevB.99.180505_mockli_2019,PhysRevB.101.014510_mockli2020} for more details. Here, we focus on the effect of magnetic impurities.

The structure of the matrix in Eq. \eqref{Eilenberger} reveals how the spin-fields couple the propagator components $\{f_0,\mathbf{f}\}$ and consequently the order-parameters $\{\psi(\mathbf{k}),\mathbf{d}(\mathbf{k})\}$. The role of the Zeeman field $B$ is to couple the singlet pairing correlations (propagators) to the triplet pairing correlations. Yet, the role of Ising-SOC is to couple the in-plane triplet correlations $f_1$ and $f_2$. The $f_3$ triplet correlations remain unaffected by the spin-fields because they are parallel to Ising-SOC and perpendicular to the Zeeman field.

We now assume an isotropic singlet state $\psi(\mathbf{k})=\psi_0$, solve Eq. \eqref{Eilenberger} for $\{\psi_0,\mathbf{d}(\mathbf{k})\}$, which together with the self-consistency condition for the order-parameters \eqref{selfconsistency} results in three sub-systems: $\{\{\psi_0,d_2(\mathbf{k})\},\{d_1(\mathbf{k})\},\{d_3(\mathbf{k})\}\}$. The solutions for $d_{1(3)}$ are the same as in Eqs. \eqref{t1} and $\eqref{t3}$, respectively. The sub-system $\{\psi_0,d_2(\mathbf{k})\}$ is coupled by the joint presence of SOC and the Zeeman field, and the pair-breaking equation is given by the characteristic equation
\begin{align}
\label{det}
\det
\begin{bmatrix}
\ln\frac{T}{T_\mathrm{cs}}+\mathcal{S}_\mathrm{s} & \mathcal{S}_\mathrm{s,t2}\\ 
\mathcal{S}_\mathrm{s,t2} & \ln\frac{T}{T_\mathrm{ct}}+\mathcal{S}_\mathrm{t2}
\end{bmatrix}
=0,
\end{align}
with $\mathcal{S}_\mathrm{s}$ defined in Eq. \eqref{Ss}, $\mathcal{S}_\mathrm{t2}$ in Eq. \eqref{t2}, and 
\begin{align}
\label{st2}
 \mathcal{S}_\mathrm{s,t2} =
\sum_{n=-\infty}^\infty 
\frac{(\pi T) B\Delta}{|\tilde{\omega}_n|B^2+(|\omega_n|+2\Gamma_m)\left[|\tilde{\omega}_n|(|\omega_n|+2\Gamma_1)+\Delta^2 \right ]}. 
\end{align}
Eq. \eqref{st2} is what couples $d_2(\mathbf{k})$ to $\psi_0$. It is only possible for $B\Delta\neq 0$.

The singlet to triplet coupling is impossible between an anisotropic singlet $\psi(\mathbf{k})$ and $d_2(\mathbf{k})$. The reason is that the product of basis function involving an anisotropic singlet state $\langle \hat{\psi}(\mathbf{k})\hat{d}_2(\mathbf{k})\rangle_\mathrm{FS}$ vanishes. Another way to interpret this is that while a finite $B\Delta$ can re-phase isotropic singlets $\psi_0$ into triplets, it necessarily de-phases anisotropic singlet Cooper pairs. This can also be understood in terms of a spin-rotation argument, which is discussed in Ref. \cite{PhysRevB.99.180505_mockli_2019}.

\begin{figure}
\centering
\includegraphics[width=0.48\textwidth]{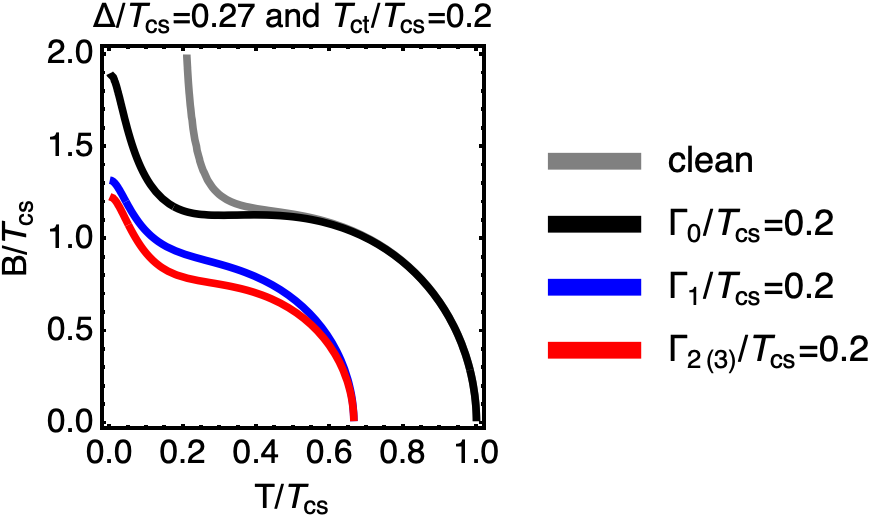}
\caption{\label{fig:5} 
Effect of impurity components on the $\{\psi_0,d_2(\mathbf{k})\}$-coupled superconducting state. The triplet components $d_{1(3)}$ are already obliterated for $\Gamma/T_\mathrm{cs} = 0.2$ and are thus not shown in the figure.
The clean (gray) curve diverges at $T=0.2 T_\mathrm{cs}=T_\mathrm{ct}$.
}
\end{figure}

Since time-reversal symmetry breaking by the magnetic field (together with SOC) couples the $\psi_0$ singlet to the $d_2$ triplet order-parameter, it is natural to ask whether time-reversal symmetry breaking by magnetic impurities could also lead to such a coupling? From the structure of the matrix in Eq. \eqref{Eilenberger} and Eq. \eqref{st2} the answer is clearly: no. 
All impurities rescale the diagonal quantities in the matrix, such that no coupling is possible. It is also clear that since the scattering rates only appear in the denominator of Eq. \eqref{st2}, the impurities suppress the coupling.
While a magnetic field can re-phase Cooper pairs, randomly oriented impurities always lead to de-phasing.

In Fig. \ref{fig:5} we plot some contours of the pair-breaking equation \eqref{det} for the parity-mixed state $\{\psi_0,d_2(\mathbf{k})\}$. 
For illustration purposes, we used the fixed value $T_\mathrm{ct}/T_\mathrm{cs}=0.2$. The divergent behavior of the clean (gray) curve reflects this value. 
For any scattering rate $\Gamma/T_\mathrm{cs}> \pi/(2e^\gamma)(T_\mathrm{ct}/T_\mathrm{cs}) \approx 0.18$, the $d_{1(3)}$ triplets would already be obliterated and are not shown in Fig. \ref{fig:5}.  
For illustration, we use a relatively small value for $\Delta/T_\mathrm{cs}=0.27$, such that the differences in the effects of $\Gamma_i$ are still visible.
The important messages of Fig. \ref{fig:5} are that now the $d_2$ triplets acquire the same robustness to impurities as the $\psi_0$ singlets; and that magnetic impurities are destructive to the singlet-triplet coupling.

\section{Discussion and conclusion\label{sec:discussion}}

We now discuss the results in the context of the literature. As examples, we comment on monolayer TMDs, superconducting thin films, heavy-fermion triplet superconductors and then conclude.

In this work, we explored ways in which the magnetic impurities can serve as a tool for probing unconventional superconductivity.  
The 2D superconductors which are of great technological and theoretical significance are anisotropic in the response to the in- and out-of-plane magnetic field. 
Also, in systems without inversion symmetry, the SOC is
anisotropic in momentum.
We studied how superconductivity in the presence of the Zeeman field and SOC can be probed by magnetic impurities that are randomly oriented and distributed across the system.
We demonstrate that the anisotropies in the distribution of the impurity spin orientation can be used as a tool to study the momentum texture of specific properties of the order parameter and its evolution with the Zeeman field.

In the absence of spin-fields such as the SOC and the Zeeman field, the effect of the scalar and magnetic impurities on the singlet superconductivity is well known.
The effect of a dilute concentration of magnetic impurities in triplet and parity-mixed superconductors remains largely unexplored.  To our knowledge, the few efforts are limited to Refs. \cite{Yavari2011a,Yavari2012a,Yavari2013a}.
Here we take into account the SOC, Zeeman field and the possible anisotropy of the superconducting order parameter.

As a specific example, we focused on the 2D Ising superconductors. 
We have addressed several relevant aspects of these systems in Refs. \cite{PhysRevB.98.144518_mockli2018,PhysRevB.99.180505_mockli_2019,PhysRevB.101.014510_mockli2020}, with special emphasis on monolayer transition metal dichalcogenides (TMDs). 
For completeness, we now comment on specifics for TMDs.
The crystal point group of the hexagonal family of monolayer TMDs is $D_{3h}$ lacking the inversion element. In the present context, the group has three relevant pairing channels \cite{PhysRevB.99.180505_mockli_2019}: an even $s$-wave $A_1'$ channel with basis function $\hat{\psi}(\mathbf{k})=1$, an odd $f$-wave $A_1'$ channel with basis function $\hat{\gamma}(\mathbf{k})\hat{\boldsymbol{z}}$, and an $if$-wave $E''$ channel with basis functions $\{\hat{\gamma}(\mathbf{k})\hat{\boldsymbol{x}},\hat{\gamma}(\mathbf{k})\hat{\boldsymbol{y}}\}$, where $\hat{\gamma}(\mathbf{k})$ is the same basis function used for Ising-SOC. Therefore, the superconducting state in monolayer TMDs is generically denoted by an $s+f+if$ state, where the $s$ refers to $\psi_0$, $f$ to $d_3$, and $if$ to $\{d_1,d_2\}$. 
According to the discussion in Sec. \ref{sec:conversion}, a small amount of impurities of any kind $\Gamma_f= \pi/(2e^\gamma)T_\mathrm{ct}$ obliterates the $f$-wave, such that we are left with the $s+if$ ($\psi_0+d_2(\mathbf{k})$) state. The $s+if$ state is obliterated by magnetic impurities $\Gamma_\mathrm{m}= \pi/(2e^\gamma)T_\mathrm{cs}>\Gamma_f$. Nonetheless, the $s+if$ state cannot be obliterated by scalar impurities $\Gamma_0$. 
In monolayer TMDs, the SOC energy scale is larger than $T_\mathrm{c}$, such that the different effects of $\Gamma_i$ are most likely insignificant.

In conventional BCS superconducting thin films, the in-plane paramagnetic critical field is of the order of a few Teslas \cite{PhysRevB.25.171_tedrow,PhysRevB.58.R2952_adams}. 
Figure \ref{fig:1}a shows that at low temperatures, the difference of the effects of $\Gamma_{1}$ and $\Gamma_{2(3)}$ should be of easy access to magnetometers. Perhaps a greater challenge is to prepare/find thin films that have magnetic impurities with preferred orientations.  Ordered magnetic impurities have been reported in superfluid $^3$He aerogels; see Ref. \cite{PhysRevLett.124.025302_halperin} and Refs. therein. It is less clear if such situations could be produced (artificially or naturally) in singlet superconductors. The situation in Fig. \ref{fig:1}b could occur in Ising systems such as thin Pb films grown on a silicon substrate \cite{PhysRevX.8.021002_liu}. The difference in the ordered magnetic impurity effects is better seen for smaller values of SOC. 
In anisotropic singlet superconductors (Fig. \ref{fig:2}), these effects are expected to be irrelevant, since the transition lines are affected equally by any kind of impurities.

To see these effects in triplet superconductors might be more challenging. To this date, to our knowledge, the only consensual triplet superconductors (besides superfluid $^3$He) are the Uranium based ferromagnets \cite{Mineev2017_review_uranium, Shimizu2019_uranium_berilium13,Ran2019_ute2}. Their critical temperature is usually below a Kelvin (and sometimes high pressures are needed), which makes impurity effects significantly harder to observe.

We addressed singlet, triplet, impurity induced crossovers from pure singlet to purely triplet, and the parity-mixed cases. In singlet superconductors, the magnetic impurities that are parallel to the in-plane magnetic field (the direction of polarization) have a weaker effect than the perpendicular impurities. In the triplet case, the in-plane components $d_{1(2)}$ are the ones that SOC suppresses. For these in-plane components, $\Gamma_{1(2)}$ has a weaker effect on $d_{2(1)}$. The $d_{1(2)}$ components are responsible for a Cooper pair polarization in the $y$ ($x$) directions. Thus, a similar rule to the singlet situation applies: \textit{the impurity component $\Gamma_i$ that is parallel to the direction of polarization has a weaker effect}. 
In the parity-mixed case, the joint action of SOC and in-plane magnetic field selects the triplet component that is polarized along $B$, namely $d_2$, to couple to $\psi_0$. 
We discussed the effect of different magnetic impurity components on the superconducting transition curves for systems with and without inversion. For the case without inversion, we specialized to the Ising-SOC type. However, these effects are general and can be generalized to other types of SOC. 
We argue that the anisotropy of the spin orientation of magnetic impurities can serve as a tool for manipulation, control, and characterization of superconducting states of both definite and mixed parity.

\begin{acknowledgments}
D.M. acknowledges the support from the Swiss National Science Foundation, Project No. 184050, and authors
acknowledge the support from the Israel Science Foundation, Grant No. 1287/15.
\end{acknowledgments}

\section*{Data availability}

The data that supports the findings of this study are available within the article.

\appendix

\section{Derivation of the expressions for the impurity self-energy}

\label{app:self_energy}

In this appendix we derive the expressions for the self energies, \eqref{Sigma0}, \eqref{Sigma1}, \eqref{Sigma2}, \eqref{Sigma3}.
\begin{figure*}
\centering
\includegraphics[width=0.7\textwidth]{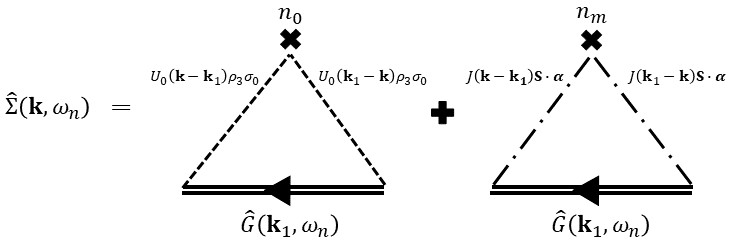}
\caption{\label{fig:A1} 
Feynman diagrams for the self-energy with scalar and magnetic
impurities proportional to $n_0$ and $n_\mathrm{m}$, respectively. 
Summation over $\mathbf{k}_{1}$ is implied. 
The contribution of the scalar impurities is given by Eq.~\eqref{Sigma0}.
Under the conditions $\overline{S_{x}S_{y}} =\overline{S_{x}S_{z}} =\overline{S_{y}S_{z}} =0$, the contribution of the magnetic impurities to the self-energy is given by the sum of the three terms, Eq~\eqref{Sigma1}, Eq.~\eqref{Sigma2} and Eq.~\eqref{Sigma3} proportional to $\overline{S_{1}^{2}}$ , $\overline{S_{2}^{2}}$
and $\overline{S_{3}^{2}}$, respectively.
}
\end{figure*}
We consider the electrons scattered off the randomly  distributed scalar and magnetic impurities.
The $4\times 4$ Gor'kov Green's function $\hat{G}\left(\mathbf{k},\omega_{n}\right)$ is related to the quasi-classical propagator via\cite{Takafumi}
\begin{align}
\hat{g}\left(\mathbf{k},\omega_{n}\right)  =\int_{-\infty}^{\infty}\frac{\mathrm{d}\xi_{\mathbf{k}}}{\pi}i\rho_{3}\sigma_{0}\,\hat{G}\left(\mathbf{k},\omega_{n}\right),
\end{align}
In this representation the magnetic exchange interaction $\mathbf{S}\cdot\boldsymbol{\sigma}$
gives rise to the term
$\mathbf{S}\cdot\boldsymbol{\alpha}$ present in the matrix of the impurity
part of the Hamiltonian presented in Eq.~\eqref{Umatix}.
The positions of impurity sites and the orientation of the impurity spins are statistically uncorrelated.
Therefore we perform the averaging over the impurity position and spin orientations independently.
This procedure leads to the the self-energy
\begin{align}
\hat{\Sigma}&
=n_{0}\underset{\mathbf{k}_{1}}{\sum}\left[U_{0}\left(\mathbf{k}-\mathbf{k}_{1}\right)\rho_{3}\sigma_{0}\right]\hat{G}\left(\mathbf{k}_{1},\omega_{n}\right)\left[U_{0}\left(\mathbf{k}_{1}-\mathbf{k}\right)\rho_{3}\sigma_{0}\right]\nonumber\\
+ & n_\mathrm{m} \sum_{\mathbf{k}_{1}}
\sum_{j=1}^3
\overline{S_{j}^{2}} \left[J\left(\mathbf{k}-\mathbf{k}_{1}\right)\alpha_{j}\right]\hat{G}\left(\mathbf{k}_{1},\omega_{n}\right)\left[J\left(\mathbf{k}_{1}-\mathbf{k}\right)\alpha_{j}\right], \label{Self_energy}
\end{align}
where $\overline{\cdots}$ denotes averaging over of all the magnetic impurities. 
In writing Eq.~\eqref{Self_energy}, we assumed that the covariance matrix of spin components is diagonal, $\overline{S_i S_j} =0$ for $i \neq j$.
The resulting self-energy, Eq.~\eqref{Self_energy} is presented in Fig.\ref{fig:A1}.

Using the relationship
\begin{align}
    &\underset{\mathbf{k}}{\sum} \approx  N_{0}\int_{-\infty}^{\infty}\mathrm{d}\xi_\mathbf{k}\int_0^{2\pi}\frac{\mathrm{d}\varphi_{\mathbf{k}}}{2\pi}
\end{align}
and the definition of the matrices $\boldsymbol{\alpha}$ which follows Eq.~\eqref{Umatix} we rewrite  \eqref{Self_energy} as the sum of four contributions, $\hat{\Sigma}=\hat{\Sigma}_0+\hat{\Sigma}_1+\hat{\Sigma}_2+\hat{\Sigma}_3$.
The contribution of the short range scalar impurities given by the first term of Eq.~\eqref{Self_energy},

\begin{align}
\hat{\Sigma}_{0} & = -i\pi n_{0}N_{0}U_{0}^{2}\rho_{0}\sigma_{0}\int\frac{\mathrm{d}\varphi_{\mathbf{k}}}{2\pi}\int\frac{\mathrm{d}\xi_\mathbf{k}}{\pi}i\rho_{3}\sigma_{0}\hat{G}\left(\mathbf{k},\omega_{n}\right)\rho_{3}\sigma_{0}\nonumber\\
 & =-i\Gamma_{0}\rho_{0}\sigma_{0}\left\langle \hat{g}\left(\mathbf{k},\omega_{n}\right)\right\rangle _{\mathrm{FS}}\rho_{3}\sigma_{0},
\end{align}
reproduces Eq.~\eqref{Sigma0} with the scattering rate $\Gamma_0$ defined in the main text.
The contribution of the $x$ impurity spin-component can be similarly obtained from the $j=1$ term in Eq.~\eqref{Self_energy},
\begin{align}\label{app:S1}
\hat{\Sigma}_{1} & =-i\pi n_\mathrm{m}N_{0}J^{2}\overline{S_{1}^{2}} \rho_{0}\sigma_{1}\int\frac{\mathrm{d}\varphi_{\mathbf{k}}}{2\pi}\int\frac{\mathrm{d}\xi_\mathbf{k}}{\pi}i\rho_{3}\sigma_{0}\hat{G}\left(\mathbf{k},\omega_{n}\right)\rho_{3}\sigma_{1}\nonumber\\
 & =-i\Gamma_{1}\rho_{0}\sigma_{1}\left\langle \hat{g}\left(\mathbf{k},\omega_{n}\right)\right\rangle _{\mathrm{FS}}\rho_{3}\sigma_{1}.
\end{align}
Equation \eqref{app:S1} coincides with \eqref{Sigma1} with the scattering rate $\Gamma_{1}$ defined in the main text.
The two remaining contribution to the self energy, \eqref{Self_energy} are similarly shown to reproduce Eqs.~\eqref{Sigma2} and \eqref{Sigma3}.

For  isotropic distribution of impurity spin orientations, 
$\overline{S_{1}^{2}}=\overline{S_{2}^{2}}=\overline{S_{3}^{2}}$, which leads to separate scattering rates to be equal, $\Gamma_{1}=\Gamma_{2}=\Gamma_{3}=\frac{1}{3}\Gamma_{m}$,
where $\Gamma_{m}$ is defined following Eq. \eqref{Sigma3}. 
In the general case of anisotropic spin distribution $\overline{ S_{1}^{2}}+\overline{ S_{2}^{2}}+\overline{S_{3}^{2}} =S(S+1)$ which leads
to the relations, $\Gamma_{1}+\Gamma_{2}+\Gamma_{3}=\Gamma_{m}$.
The latter serves as a constraint on a separate scattering rates imposed by the magnitude of impurity spin being constant.

\nocite{*}
\bibliography{biblio}

\end{document}